\documentclass[aps,prb,twocolumn,showpacs,preprintnumbers,superscriptaddress,amsmath,amssymb,longbibliography]{revtex4-2}

\usepackage{dcolumn}
\usepackage{bm}
\usepackage[normalem]{ulem}    
\usepackage{xcolor}
\usepackage{textcomp,gensymb}   
\usepackage{tikz}
\usepackage{physics}
\usepackage[pdftex,colorlinks=true,pdfstartview=FitV,linkcolor=blue,citecolor=blue,urlcolor=blue]{hyperref}
\usepackage{tabu}
\usepackage{soul}

\begin{document}


\title{Surface reconstruction-driven band folding and spin–orbit enhancement at the $\alpha$-antimonene/Au(111) interface}

\author{Thomas Pierron}
\affiliation{Laboratoire de Physique et d’Étude des Matériaux (LPEM), ESPCI Paris, PSL University, CNRS UMR8213, Sorbonne University, 75005 Paris, France}

\author{Jos\'{e} de Jes\`{u}s Villalobos Castro}
\affiliation{Universit\'{e} Paris-Saclay, ONERA, CNRS, Laboratoire d'Etude des Microstructures (LEM), F-92322 Ch\^{a}tillon, France}

\author{Etienne Barre}
\affiliation{Laboratoire de Physique et d’Étude des Matériaux (LPEM), ESPCI Paris, PSL University, CNRS UMR8213, Sorbonne University, 75005 Paris, France}

\author{Dan Wang}
\affiliation{ Laboratoire de Physique et d’Étude des Matériaux (LPEM), ESPCI Paris, PSL University, CNRS UMR8213, Sorbonne University, 75005 Paris, France}

\author{Stephane Pons}
\affiliation{Laboratoire de Physique et d’Étude des Matériaux (LPEM), ESPCI Paris, PSL University, CNRS UMR8213, Sorbonne University, 75005 Paris, France}

\author{Dimitri Roditchev}
\affiliation{Laboratoire de Physique et d’Étude des Matériaux (LPEM), ESPCI Paris, PSL University, CNRS UMR8213, Sorbonne University, 75005 Paris, France}

\author{Azzedine Bendounan}
\affiliation{TEMPO Beamline, Synchrotron Soleil, L’Orme des Merisiers Saint-Aubin, B.P.48, Gif-sur-Yvette Cedex, 91192, France}

\author{Valerie Guisset}
\affiliation{Institut N\'{E}EL, CNRS and Universit\'{e} Joseph Fourier, BP166, F-38042 Grenoble Cedex 9, France}

\author{Philippe David}
\affiliation{Institut N\'{E}EL, CNRS and Universit\'{e} Joseph Fourier, BP166, F-38042 Grenoble Cedex 9, France}

\author{Johann Coraux}
\affiliation{Institut N\'{E}EL, CNRS and Universit\'{e} Joseph Fourier, BP166, F-38042 Grenoble Cedex 9, France}

\author{Lorenzo Sponza}
\email{lorenzo.sponza@onera.fr}
\affiliation{Universit\'{e} Paris-Saclay, ONERA, CNRS, Laboratoire d'Etude des Microstructures (LEM), F-92322 Ch\^{a}tillon, France}
\affiliation{European Theoretical Spectroscopy Facility (ETSF), 17 Sart-Tilman B-4000 Li\`{e}ge, Belgium}

\author{Sergio Vlaic}
\email{sergio.vlaic@espci.fr}
\affiliation{Laboratoire de Physique et d’Étude des Matériaux (LPEM), ESPCI Paris, PSL University, CNRS UMR8213, Sorbonne University, 75005 Paris, France}


\begin{abstract}
The electronic properties of the two-dimensional (2D) $\alpha$ phase of antimonene are unique, featuring unpinned Dirac cones that can be moved with strain. Here we investigate the structural and electronic properties of an epitaxial 2D $\alpha$-antimonene, grown on Au(111). Using angle-resolved photoemission spectroscopy and density-functional theory, we reveal a strong hybridization at the Sb/Au interface, which imprints a rectangular reconstruction in the Au states, producing a band folding and hybrid bands exhibiting trigonal pockets. Additionally, hybridization displaces part of the Au wavefunction in regions of large electrostatic potential gradient, thereby enhancing spin-orbit splitting. Our work underscores that the pristine electronic properties of $\alpha$-antimonene may be deeply modified by its substrate, and even overwhelmed by the bands of the latter, and also shows that spin-orbit interaction in a heavy metal (Au) can be substantially enhanced by a lighter element (Sb).

\end{abstract}


\maketitle

\section{Introduction}

Among the new two-dimensional (2D) materials recently discovered, antimonene is receiving an increasing attention due to its intriguing structural and electronic properties~\cite{Ares2018,DongreS2022,Carrasco2023,Li2025}. 
It exists in two stable phases: the puckered $\alpha$-phase, structurally analogous to phosphorene, and the corrugated honeycomb $\beta$-phase~\cite{Wang2015b}.
Owing to their buckled geometry, both allotropes can sustain large strain, making it an effective tuning parameter controlling their electronic dispersion~\cite{Zhang2015b}. Indeed, strain-driven non trivial band topology~\cite{Chuang2013,Huang2014,Zhao2015} and the presence of strain controllable unpinned Dirac cones~\cite{Lu2016a,Lu2022} have been predicted for the $\beta$ and $\alpha$ phases respectively. 

These theoretical predictions where mainly made for freestanding antimonene. In realistic systems, the 2D layer's electronic properties can be profoundly modified by the interaction with the substrate, in which case the 2D layer cannot be considered free-standing. Actually, for many 2D materials, hybridization with a metallic substrate generates interface-driven electronic states with no analogue in the freestanding layer. Well-known examples include graphene on transition metals, where hybridization can suppress the Dirac cone \cite{Sutter2009,Wang2021} or  induce spin-polarization \cite{Usachov2015,Vincent2020}. Similar effects occur in transition metal dichalcogenides supported on noble metals where a strong renormalization of the valence band has been observed for MoS$_{2}$ on Au(111) \cite{Miwa2015,Bruix2016} and to a contact-induced semiconductor-to-metal transition in WS$_{2}$ \cite{Dendzik2017}. As far as $\alpha$-antimonene is concerned, it remains to be established whether the characteristic unpinned Dirac cones, which have been reported in the case of SnS(001) and SnSe(001) substrates~\cite{Lu2022,Shi2020,Shi2020b}, survive on other kinds of supports.


Here we address this question for the case of Au(111) susbtrate by combining angle-resolved photoemission spectroscopy (ARPES), and density-functional theory (DFT). We find that, despite a minimal out-of-plane relaxation of the top Au layers, the strong Sb–Au interaction enforces the rectangular surface periodicity of $\alpha$-antimonene on the Au(111) electronic states. This symmetry mismatch produces a folding of Au bands into the antimonene Brillouin zone, where they acquire measurable spectral weight and hybridize with Sb states. The resulting mixed Sb–Au bands form characteristic trigonal patterns set by the three rotational domains and display a significant enhancement of spin–orbit splitting. These results establish $\alpha$-antimonene/Au(111) as a model platform to harness  surface reconstruction-driven band folding and hybridization-enhanced spin–orbit physics at 2D/metal interfaces.


\section{Methods}
All the experiments have been conducted in ultra-high vacuum conditions with a base pressure below 2$\cross$ 10$^{-10}$mbar. The Au(111) substrate has been cleaned by cycles of Ar sputtering and annealing at 550$^\circ$C. Antimony has been deposited at room temperature followed by a mild annealing at 80$^\circ$C. Angle resolved photoemission electron spectroscopy (ARPES) data have been recorded with linealry s-polarizerd light at 21.21 eV at 20 K. Scanning tunneling microscopy (STM) images and low energy electron diffraction (LEED) patterns have been recorded at room temperature.  

\textit{Ab-initio} simulations have been carried out in the framework of DFT using the Quantum ESPRESSO package~\cite{Giannozzi2009,Giannozzi2017}.
The PBEsol~\cite{Perdew2008} exchange correlation potential with Grimme-D2~\cite{Grimme2006} van der Waals interactions has been chosen in all calculations.
Ultrasoft pseudopotentials with core corrections have been employed setting cutoff energies to 45~Ry and 540~Ry for wavefunctions and density respectively.
The Brillouin zone of orthorombic supercells has been sampled with a $8\times5\times1$ Monkhorst-Pack grid while a  $9\times9\times1$ Monkhorst-Pack has been used for the hexagonal cell of clean gold. 
In all cases, we have used simulation supercells comprising about 20 \AA{} of empty space in the vertical direction to avoid artefact interactions between replicas of the system.
Structural optimizations have been performed on selected atoms (see text) neglecting spin-orbit interaction and using the Broyden-Fletcher-Goldfarb-Shanno algorithm stopped when all force components on all atoms were lower than 10$^{-4}$ Ry/Bohr and total energy differences lower than 10$^{-6}$~Ry.

\section{Growth of antimonene on Au(111)}
The free-standing $\alpha$-antimonene monolayer crystallizes in a puckered rectangular lattice composed of two vertically displaced sublayers, analogous to phosphorene. Its anisotropic structure defines two inequivalent in-plane directions: the zig-zag direction, running along the ridges of the puckered network, and the armchair direction, perpendicular to them. When grown on (111)-noble metals surfaces, this structure undergoes changes that have been addressed by several experimental and theoretical studies~\cite{Shao2018,Mao2018,Zhu2019,Niu2019,Sun2020,Liu2022}. Recently, it has been unveiled that on Cu(111) and Ag(111), antimony initially grows as a surface alloy with the underlying metal and, after a critical coverage corresponding to the alloy saturation at the metal surface, antimony assumes a rectangular puckered structure~\cite{Zhang2022a}, modified with respect to the free-standing case. The two atomic planes have a vertical separation comparable to the in-plane interatomic distance. Along the zig-zag direction, antimony assumes a doubled periodicity with respect to free standing antimonene to match the lattice constant of the underlying substrate resulting in a stretched $2 \times 1$ reconstruction of $\alpha$-antimonene. This crystal structure correspond to a $(3 \times \sqrt{3})$ reconstruction with respect the metallic substrate. The growth of antimony on Au(111) is even more complicated. Cantero and co-workers proposed a model where, after the formation of the SbAu$_{2}$ surface alloy in the initial stages of the growth, for increasing coverage the alloy dissolves and its Sb atoms participate to the growth of a flat, single atomic layer on top of a clean Au(111) surface~\cite{Cantero2021}. Already the first atomic Sb layer assumes a $(3 \times \sqrt{3})$ surface structure with respect to the $1 \times 1$ of the underlying substrate. The deposition of ab additional Sb monolayer results in the formation of the puckered structure with the same in-plane reconstruction with respect to the substrate, as in the case of Cu(111) and Ag(111), without the presence of the Sb-Au alloy. 

\begin{figure}
 \includegraphics[width=0.98\linewidth]{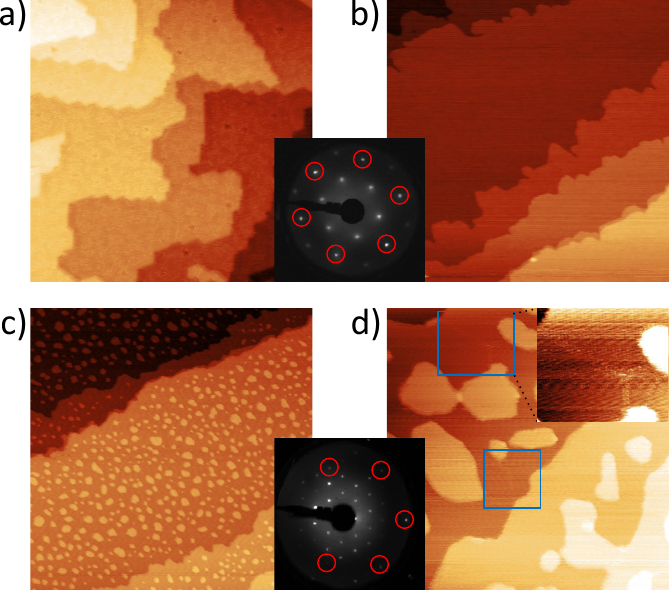}
  \caption{STM images ($V = -1$ V, $I =1$ nA of Sb/Au(111) surfaces acquired at different antimony coverage: a) 0.15 ML, b) 0.6 ML, c) 0.8 ML, d) 1.4 ML. Image size is 100 $\times$ 100 nm$^2$. Representative LEED pattern acquired at 75 eV are shown with the $1 \times 1$ Au diffraction spots highlighted by red circles. Blue rectangles in d) highlight the position of grain boundaries between the three possible rotational domains. Inset in d) shows the region highlighted with the top blue rectangle, with enhanced contrast for a better identification the grain boundary.}
  \label{fgr:example}
\end{figure}

These phases of the growth of Sb on Au(111) can be identified by STM and LEED measurements shown in Fig.~\ref{fgr:example}. 
The initial deposition of Sb on Au(111) results in slightly corrugated atomic terraces up to a coverage of about 0.15 ML (Fig.~\ref{fgr:example}a). Increasing the coverage up to 0.6 ML makes the terraces flatter, with no discernible 2D island or cluster (Fig.~\ref{fgr:example}b). In these two regimes the LEED patterns show the typical $(\sqrt{3} \times \sqrt{3})$ reconstruction that has been previously attributed to the antimony-metal alloy in the case of Au~\cite{Cantero2021}, Ag and Cu~\cite{Zhang2022a} substrates. A slight increase of the deposited amount (0.8 ML) result in the formation of atomically thin islands with high nucleation density and average size of the order of few nanometers (Fig.~\ref{fgr:example}c). Above 1 ML deposition the islands' density decreases while their size increases (Fig.~\ref{fgr:example}d). In these two cases, the LEED patterns reveal a $(3 \times \sqrt{3})$ reconstruction already reported by Cantero and co-workers~\cite{Cantero2021}, corresponding to the rectangular unit cell discussed above. Due to the three-fold symmetry of the underlying substrate, three rotational domains of the rectangular unit cell coexist on the surface. Indeed, in Fig.~\ref{fgr:example}d, two domains boundaries in the first antimony layer, with the typical 120$^\circ$ symmetry, are highlighted by blue rectangles. Inset in Fig.~\ref{fgr:example}d highlights the top grain boundary. It is important to notice that the structural transition occurs at 0.8 ML. The nanostructures imaged in Fig.~\ref{fgr:example}c are typically smaller than the LEED coherence length ($\approx 10$~nm) and cannot be responsible for the measured change in the diffraction pattern which we therefore ascribe to the underlying layer, not composed anymore of a SbAu alloy.

\section{Electronic properties of antimonene on Au(111)}

\begin{figure}
 \includegraphics[width=0.98\linewidth]{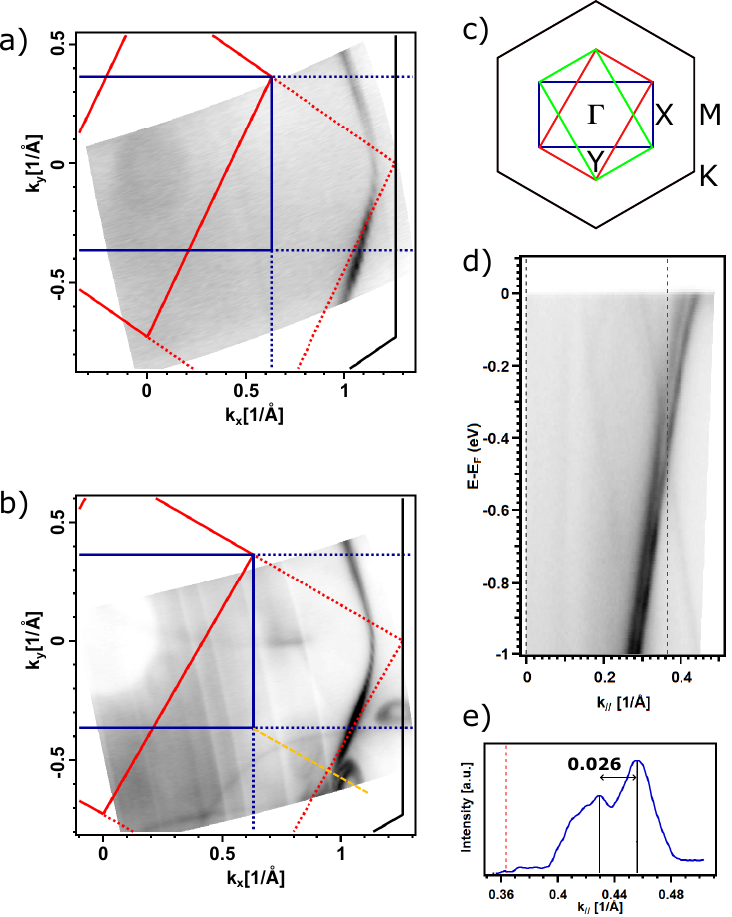}
  \caption{Fermi surfaces of the clean Au(111) a) and $\alpha$-antimonene on Au(111) b). 
  c) Schematic representation of the Au(111) Brillouin zone (black) and the three rotational domains of the $(3 \times \sqrt{3})$ rectangular $\alpha$-antimonene overlayer. High symmetry points are indicated for the Au(111) surface and the Sb domain in blue color. 
  d) Electronic dispersion along the yellow line in b) showing spin-split bands originating from Au atoms. e) Momentum distribution curve at the Fermi level of the bands in d highlighting the splitting of the bands.}
  \label{fgr:arpes}
\end{figure}

From now on we focus on the electronic properties of $\alpha$-antimonene on Au(111). Figure~\ref{fgr:arpes} compares the Fermi surfaces of clean Au(111) (Fig.~\ref{fgr:arpes}a) with the one after the growth of $\alpha$-antimonene (Fig.~\ref{fgr:arpes}b) evidencing the presence of new electronic states with trigonal symmetry. To help identification of the position of these new bands in reciprocal space, the Brillouin zones (BZ) are drawn in Fig.~\ref{fgr:arpes}c), where the Au(111) BZ is sketched in black while the Sb BZs are represented in blue, red and green for the three possible rotational domains of the $(3 \times \sqrt{3})$ reconstruction. The high symmetry points are indicated only for Au(111) (M and K) and the blue Sb BZs (X and Y). In Figs.~\ref{fgr:arpes}a) and~\ref{fgr:arpes}b), the Au and two of the antimonene rotational domains BZs are sketched on top of the Fermi surfaces where the dotted lines corresponds to second BZs of a given domain. 
In Fig.~\ref{fgr:arpes}b), due to strong ARPES matrix elements~\cite{Moser2017}, new trigonal features are visible along the $\Gamma$ - Y direction, close to the Y point in the first, second and third BZs of the red domain along with the second BZ of the blue domain. These new electronic states show a clear splitting in reciprocal space, best visible in Fig.~\ref{fgr:arpes}d). There, the electronic dispersion is shown along the orange dashed line drawn in the Fermi surface of  antimonene on Au(111). Two almost linear bands are visible, crossing the Y point, with the apex above the Fermi level. Their energy separation seems to slightly depend on the energy increasing towards the Fermi level with a value of about 0.026 \AA$^{-1}$ (inset of Fig.~\ref{fgr:arpes}d).

Similar triangular features have been reported for silicene on Ag(111), where the observed bands were interpreted either as Umklapp replicas of Ag bulk states or as silicene-derived Dirac cones modified by the substrate ~\cite{Mahatha2014,Sheverdyaeva2017,Feng2019}. This analogy naturally raises the question of whether the trigonal bands observed here originate from a similar mechanism. As shown below, neither a simple Umklapp process nor Dirac-like dispersions can account for our data.

\begin{figure}
 \includegraphics[width=0.98\linewidth]{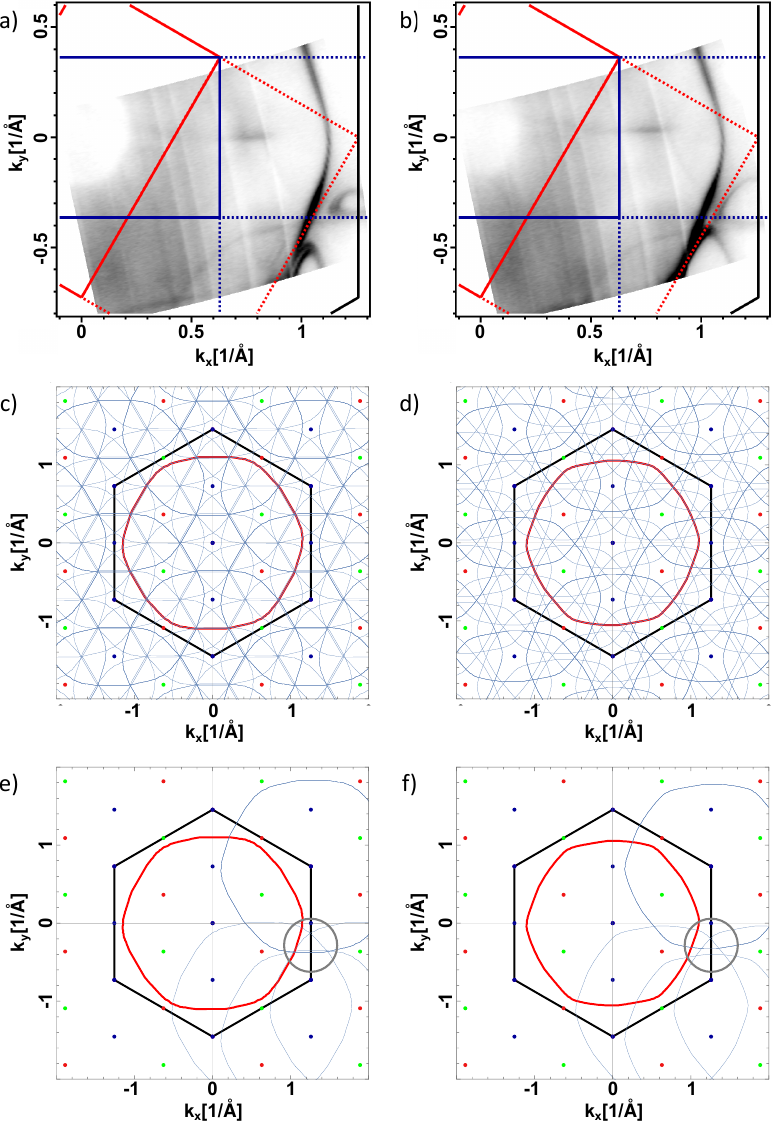}
  \caption{Constant energy maps at -0.1 eV (a) and -0.4 eV (b). The BZs are indicated in the same manner as is Fig. 2. c) and d) shows the theoretical Umklapp scattering constant energy maps at the same energies as discussed in the text. The blue, red and green dots indicate the reciprocal space vectors of the three $\alpha$-antimonene rotational domains. The black solid hexagon highlight the Au(111) BZ and the red roundish hexagon represents the symmetrized Au(111) electronic states derived for clean Au(111). Thin blue lines corresppond to the Umklapp pattern. e) and f) shows only the scattered gold band that produces the trigonal features.}
  \label{fgr:brehmstralung}
\end{figure}

The first step in understanding the origin of the trigonal features is to examine whether they could arise from a standard surface Umklapp process, in which photoemitted electrons of the substrate exchange a reciprocal lattice vector of the $\alpha$-antimonene $(3 \times \sqrt{3})$ reconstruction~\cite{Anderson1976,Westphal1983,Mugarza2002,Shikin2003,Shikin2004,Bengi2012}.
Figures~\ref{fgr:brehmstralung}a) and~\ref{fgr:brehmstralung}b) show experimental constant-energy maps at $-0.1$~eV and $-0.4$~eV,  respectively, where the Au(111) contours display their expected hexagonal symmetry, while the trigonal features appear within the second and third Brillouin zones of the antimonene domains.
To test the Umklapp scenario, we generated (theoretical) constant-energy maps by simply replicating the symmetrized Au(111) bands at all reciprocal-lattice vectors of the reconstruction for the three rotational domains. The thus-generated patterns are displayed in Figs.~\ref{fgr:brehmstralung}c) and~\ref{fgr:brehmstralung}d), while Figs.~\ref{fgr:brehmstralung}e) and~\ref{fgr:brehmstralung}f) report only the relevant replicated bands whose intersection forms trigonal shapes in the same region as those observed in the experiment. At first sight the Umklapp process reproduces the experimentally observed patterns but a closer inspection reveals that the calculated behavior differs from the experimental one in two essential aspects.

First, the simulated Umklapp triangles shrink with increasing binding energy, whereas the experimental triangles in Figs.~\ref{fgr:brehmstralung}a) and~\ref{fgr:brehmstralung}b) expand, i.e. they display the opposite trend. Second, the Umklapp replicas never exhibit the momentum-space splitting on all three sides that characterizes the experimental features. This is particularly evident when comparing the experimental triangle inside the (blue) second antimonene Brillouin zone with its simulated counterpart in Figs.~\ref{fgr:brehmstralung}e) and~\ref{fgr:brehmstralung}f). These discrepancies demonstrate that a conventional surface Umklapp mechanism cannot account for the trigonal features, indicating that an additional mechanism, beyond simple band replication, is required.

\begin{figure*}
  \includegraphics[width=1.00\linewidth]{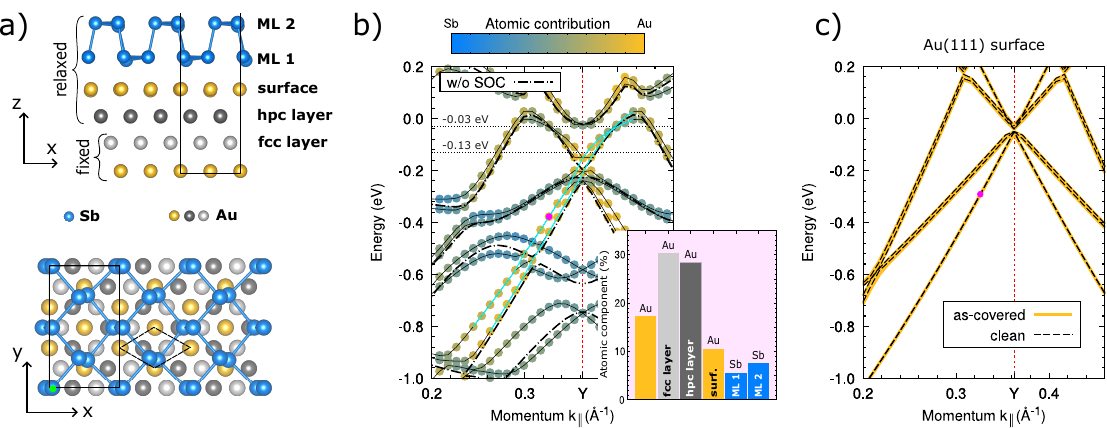}
  \caption{a) Ball-and-stick representation of the simulated Sb/Au(111) system from the side view (left) and from above (right) made with the VESTA free software~\cite{vesta}. The black rectangle defines the orthorhombic simulation cell, the black diamond draws the hexagonal unitary cell of a clean Au(111) surface. Additional information is given on the relaxation strategy and on layer labels. The axis passing through the green spot is the abscissa of Fig.~\ref{fgr:wfc_and_v}. b) DFT band structure of the Sb/Au(111) along $\Gamma-Y$ across the $Y$ point. The color code indicates the atomic character of the electronic state, yellow standing for pure Au states and blue for pure Sb states. The structure without spin-orbit coupling is also reported for comparison (black dash-dotted). Bands highlighted in cyan correspond to the measurements of Fig.~\ref{fgr:arpes}d. Horizontal dotted lines at -0.03~eV and -0.13~eV correspond to the cuts at which the maps of Figs.~\ref{fgr:trigonals}c and ~\ref{fgr:trigonals}d 
  have been taken. A magenta spot highlights a notable electronic state whose atomic-layer contributions are reported in the inset histogram.
  c) Band structure of the Au(111) surface in the reconstructed supercell in a fully relaxed calculation (\emph{clean}, dashed black) and in a calculation with atoms in the same position as in the Sb/Au(111) system (\emph{as-covered}, yellow solid). The magenta spot has the same meaning as in panel b. }
  \label{fgr:AuSb_fatbands}
\end{figure*}

\begin{figure*}
    \centering
    \includegraphics[width=\linewidth]{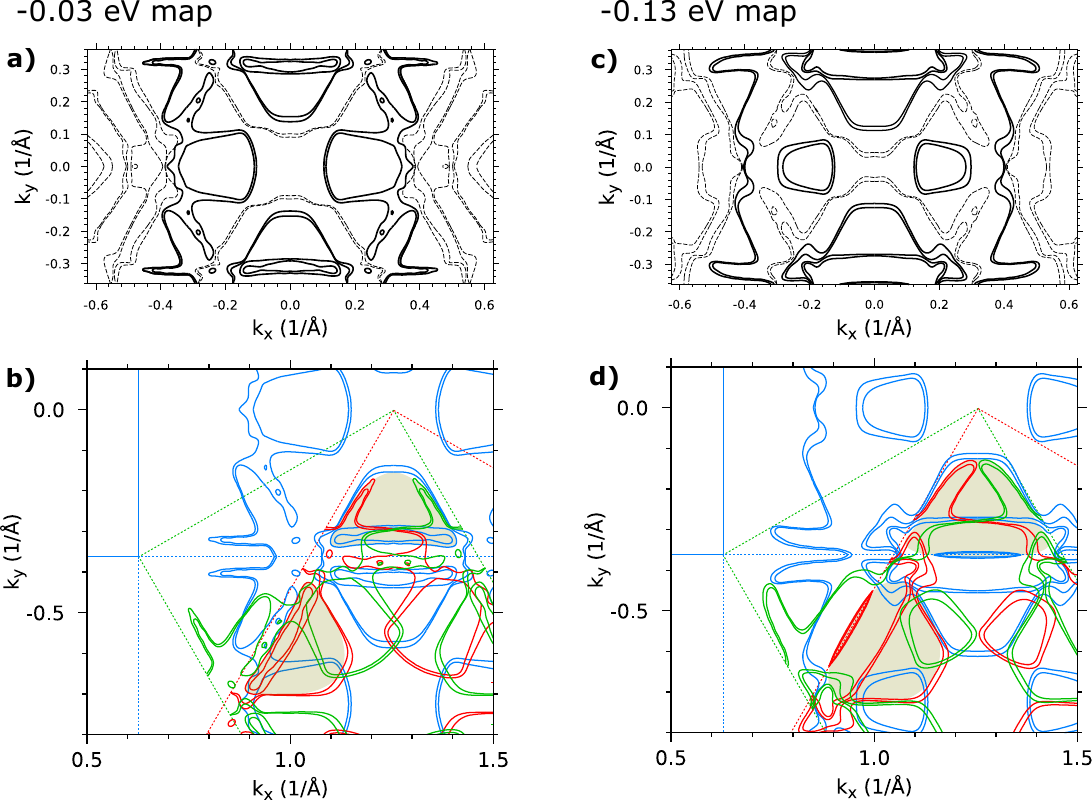}
    \caption{a) Isoenergy map from DFT calculation extracted at -0.03~eV in the first Brillouin zone. Solid contours are relevant for the formation of trigonal features. b) Identification of the trigonal features (shaded areas) of Fig.~\ref{fgr:brehmstralung}a arising outside the first Brillouin zone as a result of the rotational domains: blue = $0^\circ$, red = $+60^\circ$ and green = $-60^\circ$). c-d) The same analysis as in a and b at -0.13~eV to be compared with Fig.~\ref{fgr:brehmstralung}b.}
    \label{fgr:trigonals}
\end{figure*}

\section{First-principle interpretation} 

To unveil the actual nature of the gold surface states we employed DFT calculations. 
In the simulation cell, the $\alpha$-Sb layer is deposited on the (111) surface of a slab formed of four layers of gold in the FCC structure. 
We used cell parameters coming from the experiment (LEED) assuming a Au(111) surface lattice parameter of 2.89~\AA which we kept fixed during relaxation, together with the atomic positions of the two deepest Au layers.
All other atomic positions have been relaxed, as summarised in Fig.~\ref{fgr:AuSb_fatbands} a).

Our simulations demonstrate that the interaction between the Sb layer and the Au(111) surface is strong enough to modify the gold surface electronic structure.
Thus, the measured features are ascribed to Au states that, as a result of the hybridization with Sb, (A) acquire spectral weight with the orthorhombic periodicity, (B) form trigonal patterns due to the rotational domains of the overlayer and (C) get their spin-orbit coupling enhanced because of a displacement of their wavefunction.

\subsection{Superperiodicity on Au bands}

In Fig.~\ref{fgr:AuSb_fatbands}b, the DFT band structure of Sb/Au(111) is reported across the zone border at $Y$ along the $\Gamma - Y$ direction.
Two bands are highlighted: we relate them to the doubly-split measured band of Fig.~\ref{fgr:arpes}d.
Small discrepancies in offset and separation from the experiment are compatible with the level of approximation of the simulations.
The color code indicates that the doubly-split band is basically made of electrons localized on Au atoms.
A more detailed decomposition into atomic-layer contributions for the state highlighted with a magenta bullet is reported in the inset histogram.
It shows that the main contributions come from the gold layers below the topmost one (surface).

To better understand the role played by Au and Sb in forming these bands, we have removed the Sb layer and analysed the band structure of the Au(111) surface inside the reconstructed supercell in two configurations: the \emph{clean} one and the \emph{as-covered} one.
The former results from folding the bands of a relaxed Au(111) surface into the orthorhombic supercell.
The latter comes from a calculation where Au atoms occupy the very same position as in the calculation of Fig.~\ref{fgr:AuSb_fatbands}.
In the \emph{as-covered} simulation, the orthorombic periodicity is the physical one because it is imposed by the orthorhombic surface reconstruction reminiscent of the Sb overlayer, although weak, while in the \emph{clean} calculation it is an arbitrary choice.
Results are summarised in Fig.~\ref{fgr:AuSb_fatbands}c.
Let us first focus on the \emph{clean} results (black dashed lines). 
A pair of linearly dispersing and almost degenerate bands can be identified easily in approximately the same energy range as those highlighted in Fig.~\ref{fgr:AuSb_fatbands}.
They have no equivalent in the unitary hexagonal cell of the Au(111) surface because they are a consequence of the band folding.
Conversely, when the ``as-deposited'' atomic positions are taken into account (solid yellow lines), the superperiodicity acquires a physical meaning and these linear states may become measurable, e.g. through ARPES.
Note, however, that the \emph{clean} and the \emph{as-deposited} band plots look very similar, indicating that what really endows them with spectral weight is the periodicity of the induced reconstruction, not the details of the atomic arrangement.

\subsection{Formation of trigonal pockets}

From Figs~\ref{fgr:brehmstralung}a and~\ref{fgr:brehmstralung}b we see that the size of the trigonal features increases with increasing binding energy, at least in isoenergy maps taken above the zone border crossing. 
To better assess the identification of the electronic states of the previous section, here we check that similar features appear in simulated data as well and display the same energy dependence.
In Figs.~\ref{fgr:trigonals}a and~\ref{fgr:trigonals}b we report the DFT isoenergy maps inside the entire first Brillouin zone of Sb/Au(111) at -0.03~eV and -0.13~eV respectively (cfr. horizontal dotted lines in Fig.~\ref{fgr:AuSb_fatbands}b). 
In choosing these energies, which actually differ from the experimental ones, we paid attention to the offset and steepness discrepancies between simulations and measurements.
Therefore we selected two energy cuts that fall below the simulated Fermi energy and above the energy at which the two gold bands cross the Brillouin zone border ($\approx -0.2$~eV).
In Figs.~\ref{fgr:trigonals}a and~\ref{fgr:trigonals}b it is already possible to recognise some trapezoidal features close to the $Y$ point, which display the expected energy behaviour, i.e. they are wider at higher binding energy.

In addition, if we take into account the three rotational domains of the orthorhombic cell and superimpose the maps, other bands intersect these trapezoids giving rise to shapes that look even more similar to the triangular patterns measured in ARPES.
The demonstration is depicted in Figs.~\ref{fgr:trigonals}c and~\ref{fgr:trigonals}d where the superposition of rotated and shifted isoenergy maps is reported in the same momentum space as in Figs.~\ref{fgr:brehmstralung}a and~\ref{fgr:brehmstralung}b. 
For the sake of clarity, in the plot not all bands of the isoenergy map are reported but only those that are relevant for the triangular shapes to arise (black solid traits in Figs.~\ref{fgr:trigonals} a) and~\ref{fgr:trigonals} b).

This analysis supports an interpretation of the observed featured as gold states, and rules out
a standard umklapp effect.
Moreover it allows us to exclude the triangular shapes as being due to conic features.

\subsection{Enhanced spin-orbit coupling and hybridization}

\begin{figure}
 \includegraphics[width=0.98\linewidth]{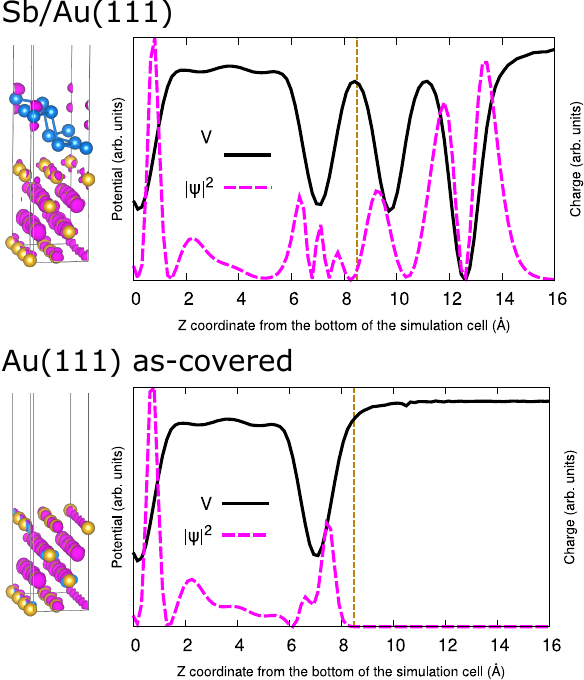}
  \caption{Left: Isosurface of $|\Psi|^2$ of the states highlighted with magenta bullets in Figs.~\ref{fgr:AuSb_fatbands}b (top) and~\ref{fgr:AuSb_fatbands}d (bottom)  made with the VESTA free software~\cite{vesta}. Right: Profile of the total potential (solid black) and $|\Psi|^2$ (dashed magenta) along a vertical axis passing through the $p_z$-like lobes of the topmost Sb atom. The brown dashed line at 8.5~\AA{ }is placed approximately halfway between the Sb monolayer and the Au surface. }
  \label{fgr:wfc_and_v}
\end{figure}

Even though the analysed bands display a dominant Au contribution, a remarkable feature with respect to the Au(111) surface states is the enhancement of the effect of SOC. 
To give an example, at -0.13~eV, the calculated SOC is negligible in the gold surface and increases to 0.008~\AA$^{-1}$ in the Sb/Au(111) calculations to be compared with the experimental 0.026~\AA$^{-1}$ (measured at the Fermi level inset of Fig.~\ref{fgr:arpes}d).
This may seem counterintuitive, because Sb, being lighter than Au, is not expected to generate SOC effects that can increase the splitting.
Below we demonstrate that the atomic number is not the most important parameter here.

The increase of the splitting as due to the interaction between Sb and Au is apparent in the band evolution reported in Fig.~\ref{fgr:hybridization_evolution} in Appendix A.
Let us now explain the origin of this effect.

We considered the states of Sb/Au(111) and the \emph{as-covered} gold surface highlighted by magenta circles in Figs.\ref{fgr:AuSb_fatbands}b and~\ref{fgr:AuSb_fatbands}c, so at about -0.4 eV, that we chose as representative.
The wavefunction squared of these states (partial charge density $|\Psi|^2$) is plotted in 3D with magenta lobes on the left side of Fig.~\ref{fgr:wfc_and_v} in the two systems. 
In the gold surface, the wavefunction is mostly localised in the hpc and fcc layers of the gold substrate (cfr. Fig.~\ref{fgr:AuSb_fatbands}a).
On the other hand, in Sb/Au(111) it also spreads into the Sb monolayer as a consequence of the hybridization and develops peculiar $p_z$-like lobes in the topmost Sb atoms.  
We further define a vertical axis $z$ passing through these lobes (highlighted with a green line in the left panels of Fig.~\ref{fgr:wfc_and_v} and a green circle in Fig.~\ref{fgr:AuSb_fatbands}a) and we calculated both the total potential and $|\Psi|^2$ of the considered states along this axis.
Results are reported in the right panels of Fig.~\ref{fgr:wfc_and_v}. 
The hybridization with Sb causes a spatial displacement of
the wavefunction maxima to a region where the potential varies strongly, increasing the sensitivity of the state to gradients of the potential and thus increasing the associated SOC.

\section{Conclusion}

In summary, we have shown that $\alpha$-antimonene grown on Au(111) exhibits an electronic structure profoundly modified by the strong Sb–Au interaction. Although the surface reconstruction of gold only involves relatively small atomic displacements, the rectangular symmetry of the antimonene overlayer imprints a superperiodicity that folds the Au surface and bulk bands into the antimonene Brillouin zone. These folded Au states acquire measurable spectral weight and, through hybridization with Sb, give rise to trigonal, spin-split features clearly visible in ARPES. First-principle calculations demonstrate that hybridization shifts part of the wavefunction into regions of steep potential, which accounts for the pronounced enhancement of the spin–orbit interaction. Our results highlight the $\alpha$-antimonene/Au(111) interface as an archetype of structurally-mismatched 2D/metal systems where hybridization can be exploited to tailor band folding and interfacial spin–orbit physics.

\begin{acknowledgments}
This research was founded, in part, by the French National Research Agency (ANR), grant numbers ANR-21-CE30-0043 (project SAGA) and ANR-21-CE09-0016 (project EXCIPLINT).
\end{acknowledgments}

\section*{Appendix A}

\begin{figure*}
\centering
\includegraphics[width=1.0\textwidth]{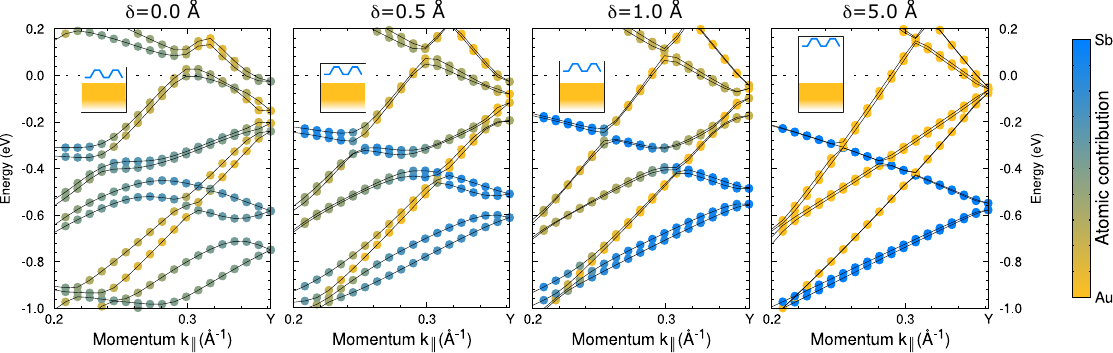}
\caption{From left to right: simulated band structure of the Sb/Au(111) system where an additional spacing $\delta$ has been introduced between the Au surface and the Sb monolayer. The $\delta=0$~\AA{ }plot corresponds to that of Fig~\ref{fgr:AuSb_fatbands}b.
The $\delta=5$~\AA{ }plot, basically corresponds to the superposition of the bands of the isolated Sb monolayer and the Au as-deposited surface.
The color-code has the same meaining as in Fig.~\ref{fgr:AuSb_fatbands}b.}
\label{fgr:hybridization_evolution}
\end{figure*}

To better appreciate the building-up of the Sb-Au hybridization, we introduced an artificial verical spacing $\delta$ in our Ab/Au(111) simulations.
By increasing $\delta$ from 0~\AA{ }to 5~\AA{ }we gradually separate the Sb monolayer from the Au surface.
Bandplots at chosen values of $\delta$ are reported in Fig.~\ref{fgr:hybridization_evolution}.
Note that no additional atomic relaxation has been performed.
We observe that the band structure evolves from the Sb/Au(111) system ($\delta=0$~\AA) to the superposition of the band structures of the Sb monolayer and the Au(111) surface ($\delta=5$~\AA). 
In particular one can easily observe that, as the Sb layer gets closer to the metal surface, the spin-orbit splitting increases almost everywhere in both the Au (yellow) and the Sb states (blue).
This clearly indicates that the enhancement of the spin-orbit coupling is a direct consequence of the Sb-Au interaction.

\end{document}